\documentstyle[twocolumn,prb,aps,psfig]{revtex}
\begin{document}
\title{Edge critical behavior at the surface transition of Ising magnets}
\author{M. Pleimling and W. Selke}
\address{Institut f\"ur Theoretische Physik B, Technische Hochschule,
D--52056 Aachen, Germany}
 
\maketitle
 
\begin{abstract}
Using Monte Carlo techniques, Ising models with ferromagnetic
nearest--neighbor interactions on a simple cubic lattice are studied.
At the surface transition, the critical exponent $\beta _2$ of
the edge magnetization is found to be non--universal, depending
on the edge and edge--surface couplings, in contrast to the situation at 
the ordinary transition. Results are compared to those 
for two--dimensional Ising magnets with chain and ladder defects.\\

\end{abstract}

\noindent {\it Pacs numbers:} 05.50+q, 68.35.Rh, 75.40.Mg\\
\vspace{1cm}

\noindent {\bf I. Introduction}
 
Critical phenomena occur not only in the bulk of a system, but also
at its surfaces and edges. To be specific, let us
consider the magnetization as the order parameter. Then, there are two
typical scenarios: (a) bulk, $m_b$, surface, $m_1$, and edge, $m_2$,
magnetizations may order at the same temperature ('ordinary
transition'), but with different power--laws, and (b) surface
and edge may order first ('surface transition') due to
strong surface couplings, followed
by ordering of the bulk magnetization ('extraordinary transition') at a lower
temperature. Surface singularities at ordinary and surface
transitions have been studied extensively for various systems, theoretically 
\cite{binder,diehl} as well as experimentally. \cite{dosch} Edge critical 
behavior, on the other hand, has attracted less attention.

Some years ago, Cardy noted and calculated the dependence of
edge critical exponents on the opening angle between the 
surfaces forming the edge, using mean field theory and
renormalization group theory of first order
in $\epsilon$ for $O(n)$ models at the ordinary
transition. \cite{cardy} Subsequent work dealt mainly with
edge criticality at the ordinary transition as well, including 
high temperature series expansions \cite{gut}, exact analyses of
two--dimensional Ising models \cite{peschel}, and applications
to polymers. \cite{bell} 

In recent Monte Carlo simulations of three-dimensional Ising models,
we confirmed and refined
the estimates obtained from renormalization group calculations  
and high temperature series expansions
on edge critical exponents at the ordinary
transition, especially on $\beta _2$ describing the vanishing
of the edge magnetization $m_2$ on approach to criticality. \cite{ps2} The 
value of $\beta _2$ varies with the opening angle, but not with
the bulk, surface and edge coupling constants which are assumed to be
ferromagnetic and of short range.

Edge critical properties at the surface transition have been
largely overlooked. Actually, we are aware of merely one
exception, where they have been mentioned briefly and   
qualitatively. \cite{larsson} However, that case deserves to
be investigated, in particular, by
Monte Carlo techniques in the framework of Ising
models, too. At the surface transition 
in three-dimensional systems, the critical fluctuations
are essentially two--dimensional, and the surface
critical exponents reflect the reduced
dimensionality. \cite{binder,diehl} On the
other hand, the edge presents a local perturbation, acting
presumably like the much studied  
chain or ladder defects \cite{fife,bariev,mccoy,ko,igloi} in two--dimensional
Ising systems. Following those
studies, intriguing non--universal edge critical behavior, depending
on the interactions at the edge, may be expected, in
contrast to the situation at the ordinary transition.  
Furthermore, previous Monte Carlo simulations on the
critical properties of Ising models at surfaces, edges, and
corners \cite{ps2,lanbi,kiel,ps1} show that present simulational
techniques, especially cluster--flip algorithms, allow to estimate reliably 
edge critical properties. Accordingly, the simulational results
may provide guidance for future analytical investigations. Finally, the
findings may encourage experimental search for 
suitable magnets or alloys \cite{dosch}, where the surface coupling
is sufficiently strongly enhanced compared to the bulk coupling.
  
The outline of the article is as follows. In the next section, the
model and its phase diagram as well as the Monte Carlo method will
be introduced. In section 3, we shall 
discuss magnetization profiles and the critical exponent
of the edge magnetization at the surface transition
in the three--dimensional Ising model for various couplings near
the edges. A brief summary concludes the paper.\\ 
 
\noindent{\bf II. Ising model with surfaces and edges}
 
We study nearest--neighbor Ising models on simple cubic lattices
with ferromagnetic interactions. The Hamiltonian may be
written in the form

\begin{eqnarray}
{\cal H} & = & -\sum\limits_{bulk} J_b S_{xyz} S_{x'y'z'}
 -\sum\limits_{surface} J_s S_{xyz} S_{x'y'z'} \nonumber \\
&& -\sum\limits_{edge-surface} J_{es} S_{xyz} S_{x'y'z'}
 -\sum\limits_{edge} J_e S_{xyz} S_{x'y'z'}
\end{eqnarray}
where the sums run over bonds between neighboring
spins, $S_{xyz}= \pm 1$, with coupling constants to be specified below.
Free boundary conditions hold for
the spins in the $yz$-- and $xz$--surface planes, while
the spins in the first and last $xy$--planes are connected by
periodic boundary conditions, see Fig.1. Thereby the
edges formed by the intersecting free surface planes are oriented
along the $z$--axis. The total number of sites or
spins is $L \times M \times N$, where $L, M$, and $N$ correspond to 
the $x, y$,and $z$--directions, respectively.
In the following, we assume $L= M$, i.e. there are $L^2$ spins
in each plane perpendicular to the edges.
The pairs of spins in the Hamiltonian (1) 
are situated either on edges with the edge
coupling $J_e$, on edge and surface sites coupled by $J_{es}$, on
surface sites with the interaction $J_s$, or with at least
one of the spins in the interior of the
system interacting with the bulk coupling $J_b$, as
depicted in Fig. 1.

The magnetization per site for lines parallel to
the $z$--axis, $m_l(x,y)$, is defined as
\begin{equation}
 m_l(x,y)= < | \sum\limits_z S_{xyz} | >/N
\end{equation}
summing over all spins of a line, with $x$ and $y$ being fixed. The brackets
denote thermal averages. The absolute value is taken to avoid
vanishing magnetizations for finite systems, as usual.
The edge magnetization, $m_2$, is then identical
to $m_l(1,1)= m_l(1,L)= m_l(L,1)= m_l(L,L)$. The surface
magnetization, $m_1$, is given
by $m_l(1, L_C)= m_l(L_C,L)= m_l(L_C,1)= m_l(L,L_C)$, where
$L_C$ refers to the one or two center lines, with $L_C= (L+1)/2$ or
$L_C= L/2 \pm 1$. The profile of the line magnetization at the
surface, $m_{ls}(i)$, is described by
\begin{equation}
 m_{ls}(i)= (m_l(x,L) + m_l(x,1) + m_l(1,y) + m_l(L,y))/4
\end{equation}
with, setting the lattice constant equal to
one, $i= x= y= 1,2,...,L$ (certainly, the profile is symmetric
around $i= L_C$). The 
bulk magnetization, $m_b$, corresponds to $m_l(L_C,L_C)$.

Of course, the line magnetizations may be affected by
finite size effects. The thermodynamic limit is
approached when $L, M, N \longrightarrow \infty$.

To elucidate bulk, surface, and edge properties of the model, we
considered a few additional quantities. In particular, we
calculated the energy of the surface spins, $E_s$, the
energy of the edge spins, $E_e$, the edge
susceptibility $\chi _{22}$ (defined as the fluctuation of the
edge magnetization \cite{cardy}), and higher moments of 
various magnetizations (Binder cumulants).

The linear dimensions, $L$ and $N$, of the $L^2 \times N$ systems ranged
from $L= 10$ to $L=80$, and from $N=10$ to $N=160$. We used the efficient 
one--cluster--flip Monte Carlo algorithm, which reduces significantly
critical slowing down.  Typically, averages were taken over a few $10^4$
clusters, after equilibration. Error bars resulted from sampling over
several realizations.

The phase diagram of the Ising model (1) is determined by the
ratio of the surface to the bulk
coupling, $r= J_s/J_b$, \cite{binder,diehl,kiel,ps1,calif} see Fig. 2. 
At $r < r_c \approx 1.50$, the model displays an ordinary transition, with 
bulk, $m_b$, and surface
magnetizations, $m_1$, ordering simultaneously at the bulk transition
temperature, $T_c= 4.5115...J/k_B$. \cite{landau,tal} The surface
critical exponents are universal, i.e. independent of the ratio $r$. For
instance, the surface magnetization $m_1$ vanishes on approach to
$T_c$ as $m_1 \propto t^{\beta _1}$, where
$t= |T- T_c|/T_c$ is
the reduced temperature, with $\beta_1= 0.80 \pm 0.01$ \cite{ps1} (this
value is robust against randomness in the surface couplings
and corrugations of the surface \cite{ps1,diehl2}). At
$r= r_c \approx 1.50$, bulk and surface still order at the same
temperature, $T_c$, but the surface critical exponents at this
'special point' are in a different universality
class, e.g., $\beta_1 \approx 0.24$. \cite{binder,diehl} 

At $J_s > r_c J_b$, the surface magnetization $m_1$ orders at the
surface transition, with the critical temperature $T_s$ being higher
than that of the bulk ordering which occurs at
the extraordinary transition, $T_c$. At
$T_s$, the surface critical fluctuations are of two--dimensional
character, and the surface critical exponents are those of the 
two--dimensional Ising model with nearest--neighbor
interactions \cite{binder,diehl}, i.e., for example, $\beta_1=1/8$.  
 
Because edges are one--dimensional, and all couplings in the
models are of short range, one expects edge quantities to become
singular at the surface transition, $T_s$, but the edge critical
exponents may differ from the corresponding surface
values \cite{ps2} (similar to the critical behavior at
one--dimensional surfaces in two-dimensional Ising
models \cite{ssli}). Most of the Monte
Carlo simulations near $T_s$ were done at fixed ratio $r= J_s/J_b= 2$, varying
then the nearest neighbor interactions along the edges, $J_e$, and
between edge and surface spins, $J_{es}$, with $J_e$ ranging from
0 to $2J_s$, and $J_{es}$ ranging from $0.5J_s$ to $2J_s$.

To determine edge critical exponents accurately, $T_s$ needs to be
determined accurately. Using standard finite--size
analyses \cite{bafi} (with the scaling variable $Nt$, implying that
the surface fluctuations are governed by the surface
correlation length which diverges like $t^{-1}$), $T_s$ may
be obtained from the location of turning points
or extrema in various thermal quantities. As illustrated and
indicated in Fig. 3, different quantities give, indeed, consistent
estimates. For $r=2$, we find $T_s= 4.9575 \pm 0.0075$. Note
that other critical temperatures of the surface transition, needed
to map the phase diagram, Fig. 2, have been obtained with less
accuracy, by doing finite--size analyses on
the location of the turning point of the
surface magnetization
for systems of moderate sizes, with up to $40^3$ sites.\\

\noindent{\bf III. Magnetization profiles and edge critical exponents}
 
The profiles of the line magnetization at the
surface, $m_{ls}(i)$, equation (3), reflect the influence
of bulk spins close to
the surfaces and edges as well as the strength of the couplings near
the edges, $J_e$ and $J_{es}$.

Typical profiles are shown in Fig. 4. At $J_e= J_{es}= J_s (=2 J_b)$, i.e., for
equal edge, edge--surface, and surface couplings, the 
effect of the bulk spins near the surfaces and edges on
the temperature dependence of the profile is illustrated
in Fig. 4a. At low temperatures, $m_{ls}(i)$ increases monotonically
with $i$, i.e. the distance from the edge. The magnetization
of the surface spins, which are connected directly to an ordered
bulk spin, is enhanced compared to the edge
magnetization, $m_2= m_{ls}(1)$. Edge spins have
the coordination number four, while
it is five for the surface spins.
In contrast, on approach to
$T_s$, the ordering of the spins falls off quickly by going from the
surface to the bulk, and the surface magnetization, $m_1= m_{ls}(L_C)$, is
pulled down below the edge magnetization by the coupling
to the bulk spin. Roughly at temperatures above $T_c$, an
interesting non--monotonic behavior in the profile shows up, with a
maximum close to the edge. It may
be explained by the fact that surface spins
next to, but on different sides of an edge, are more strongly connected
to each other, through the same neighboring bulk
spin, than spins on the flat part
of the surface with the separation distance of two. 

The profile approaches the surface magnetization $m_1$ nearly
exponentially already at rather small distances from
the maximum, i.e. $|m_{ls}(i)- m_1| \propto \exp (-ai)$; $a$ is
expected to be, as $T \longrightarrow T_s$, the inverse
correlation length. \cite{binder,diehl,ps2}

In Fig. 4b, the influence of the edge 
and edge--surface couplings, $J_e$ and $J_{es}$, on the profile, at
fixed temperature, is illustrated. The trends may be readily
understood. By increasing the coupling, $J_e$, along an edge, the
edge magnetization will rise. Likewise, a larger
interaction, $J_{es}$, of the edge spins with the neighboring surface
spins will lead to an increase of the edge magnetization. In that
case, the non--monotonic shape of the profile close to the
edge sets in at lower temperatures.

Near the surface transition, where the critical fluctuations are
of two--dimensional character, the  
edge acts like a defect line in
an essentially two--dimensional bulk Ising model; $J_e$ corresponds
to a 'chain defect' and $J_{es}$ to
a 'ladder-type defect'. \cite{igloi} The change in the topology
at the edge compared to the surface amounts to
a complicated ladder-type defect.

This analogy will be crucial in
explaining the simulational findings on
the critical exponent $\beta_2$ of the edge magnetization, defined by
\begin{equation}
 m_2 \propto t^{\beta _2}
\end{equation}
where $t$ is the reduced temperature of the surface
transition, $t= |T_s -T|/T_s$, $t \longrightarrow 0$.
To estimate $\beta_2$, we consider the effective exponent \cite{ps2,ps1}
\begin{equation}
 \beta_{eff} = d \ln m_2/ d \ln t
\end{equation}
In analysing Monte Carlo data, the derivative is replaced by a
difference at discrete temperatures. $\beta_{eff}$ approaches
$\beta_2$ as $t \longrightarrow 0$, provided finite--size effects
can be neglected.
 
To check whether the edge magnetization $m_2$ is affected by the
finite size of the Monte Carlo system, one may proceed as follows.
First, one chooses, at fixed temperature, the linear dimensions
$L$ and $N$ to be sufficiently large to reproduce the numerically
to a high degree of accuracy known thermodynamic
bulk, $m_b= m_l(L_C,L_C)$, and surface, $m_1= m_{ls}(L_C)$, magnetizations. In
a second step, $N$ may be enlarged to search for further size dependences
in $m_2$. \cite{ps2}
 
In Fig. 5, the effective exponent $\beta_{eff}$ of the edge magnetization
is displayed for various system sizes, in the
case of equal surface and edge
couplings $J_e= J_{es}= J_s= 2 J_b$. Obviously, $\beta_{eff}$ increases on
lowering the reduced temperature before it acquires, at $t< 0.15$, nearly
a plateau. Finally, on further approach to $T_s$, it decreases to zero due to
a non-vanishing edge magnetization at the critical point
in finite systems. Monte Carlo data which are largely free of
finite-size dependences have been obtained down
to $t \approx 0.06$, using systems with up to $80^3$ spins. Assuming 
the plateau--like behavior, $0.06< t < 0.15$, to hold
even closer to $T_s$ (which would
imply that corrections to scaling, i.e. to the
asymptotic power--law, equation (4), are small), we
arrive at the estimate of the critical exponent $\beta_2$ for this
set of coupling constants, $\beta_2= 0.095 \pm 0.005$. Error bars in Fig. 5
include both sample averaging and the uncertainty
in $T_s$, see above. Here and in the following, the error bar for the
value of $\beta_2$ is rather subjective, based on 'reasonable'
extrapolation of the data, which are close
to those in the thermodynamic limit, to $t \longrightarrow 0$. 

The critical exponent $\beta_2$ is significantly
lower than the critical exponent
of the surface magnetization, $\beta_1= 1/8$, reflecting the fact that
the edge magnetization is, on approach to $T_s$, larger than the
surface magnetization. We may relate this finding to results for
two--dimensional nearest--neighbor Ising model with a
ladder defect. In that model, the couplings are
identical, say, $J> 0$, throughout the system with the exception of    
the couplings, $J_l$, of the spins in one row to the spins in
a neighboring row. The ladder rows correspond to 
the edge, and the ladder couplings, $J_l$, reflect not only the
edge--surface interactions $J_{es}$, but also 
the change in topology (or connectedness to bulk spins) at
the edge compared to the surface. For the
two--dimensional Ising model with
a ladder defect, the critical exponent $\beta_l$ of
the magnetization in the ladder rows has
been determined exactly \cite{bariev,mccoy},
\begin{equation}
 \beta_l = 2 \arctan^2(\kappa_l^{-1})/\pi^2
\end{equation}
with $\kappa_l= \tanh (J_l/(k_B T_{2d}))/ \tanh (J/(k_B T_{2d}))$, where
$T_{2d}$ is the transition temperature. Thence, the critical 
exponent of the ladder row magnetization varies monotonically with the
strength of the coupling $J_l$, being smaller than 1/8
when $J_l> J$. Comparing the critical exponents $\beta_l$ and $\beta_2$, one
may attribute an effective ladder coupling
$J_l$ to the edge. Our case, $J_e= J_{es}= J_s= (2J_b)$, corresponds
to $J_l> J (=J_s)$. Of course, the
non--monotonic profiles of the line magnetization suggest that a
closer analogy between edge properties and descriptions by
two--dimensional Ising models with defect lines 
would require more complicated, extended ladder--type defects, which 
have not been treated analytically so far. \cite{igloi}

At any rate, exact results on two--dimensional Ising models with
defect lines of different types \cite{igloi,bariev,mccoy} provide
a useful framework to discuss our
simulational findings on the dependence of $\beta_2$ on the
edge and edge--surface couplings, $J_e$ and $J_{es}$. Obviously, $J_{es}$
is intimately related to $J_l$, while changing $J_e$ may be interpreted as
modifying nearest--neighbor couplings, $J_{ch}$, along a single defect
line or chain in
the two--dimensional Ising model. For
a chain defect, the critical exponent, $\beta_{ch}$, of
the magnetization 
in the defect chain varies continuously as \cite{bariev,mccoy}
\begin{equation}
 \beta_{ch} = 2 \arctan^2(\kappa_{ch})/\pi^2
\end{equation}
with $\kappa_{ch}= \exp[-2(J_{ch}/(k_B T_{2d})-J/(k_B T_{2d}))]$. Again, the
critical exponent changes monotonically, being
larger than 1/8 when $J_{ch}< J$.

Pertinent results on the edge critical exponent $\beta_2$ are
shown in Figs. 6 and 7. In Fig. 6, the effect of weakening the  
interactions between the edge and the surface, $J_{es}$, is
demonstrated. As expected from equation (6), lowering that 
coupling leads to an increase in the value
of $\beta_2$. Only Monte Carlo data are depicted which seem to be unaffected
by finite--size dependences, simulating systems with up to
$80^3$ sites. As before, the
effective exponent $\beta_{eff}$ exhibits a broad plateau
in $t$, which may allow one to extrapolate the asymptotic exponent
quite accurately. Our estimates, for $J_e= J_s= 2 J_b$, are
$\beta_2= 0.176 \pm 0.005$ at $J_{es}= J_e/2$ and
$\beta_2= 0.244 \pm 0.005$ at $J_{es}= J_e/4$.
  
Similarly, weakening of the edge interaction $J_e$ tends to 
increase the critical exponent of the edge magnetization, see
equation (7). In Fig. 7, the effective exponent $\beta_{eff}$
is shown, using again only those simulational data for the
edge magnetization which
approximate closely the infinite system. In this case, the effective
exponent depends on approach to $T_s$ nearly linearly on the
reduced temperature $t$, signalling rather strong corrections to
scaling for $m_2$. By extrapolating the findings for $\beta_{eff}$ to
the critical point, $t \longrightarrow 0$, we estimate, for
$J_{es}= J_s= 2J_b$, $\beta_2= 0.170 \pm 0.005$ in the
limiting case $J_e= 0$, and $\beta_2= 0.127 \pm 0.005$ for
$J_e= J_{es}/2$. The close agreement with the value of
the surface critical exponent $\beta_1= 1/8$, in 
the latter case, is rather fortuitous, due to
a compensation of a reduction in $\beta_2$ following from
the increase in the effective ladder coupling
stemming from the edge topology (as discussed above), and of an   
enhancement in $\beta_2$ following from the decrease in the
strength of the chain coupling.

The non--universal character of the edge transition driven
by the surface transition is also confirmed by our findings
for the critical properties of the edge
susceptibility $\chi_{22}$. However, because
the estimates are quite rough, we refrain from quoting specific
values.

The Binder cumulant at $T_s$ for the edge magnetization depends
on the edge and edge--surface couplings as well. However, this
may not be interpreted
as additional evidence for non--universality, because that cumulant
changes with, e.g., the ratio of the surface
to bulk interactions $r= J_s/J_b$ even at
the ordinary transition, while the critical exponents are universal
there. \cite{lanbi}\\

\noindent {\bf IV. Summary}
 
We studied magnetization profiles and the
critical exponent $\beta_2$ of the edge magnetization of three--dimensional
nearest--neighbor Ising models near the surface transition, $T_s$ (with
the surface coupling $J_s$ being twice as large as the bulk coupling). We
did large--scale simulations, using the single--cluster--flip Monte
Carlo algorithm. In particular,
the influence of the edge, $J_e$, and edge--surface, $J_{es}$,
couplings has been investigated.

The profiles of the line magnetization at the surface
suggest that the edge acts like a rather
complicated, extended ladder-type defect in an essentially
two--dimensional Ising magnet. The reduced dimensionality
is due to the two--dimensional character of the critical
fluctuations at the surface transition.

The edge critical exponent $\beta_2$ varies
continuously with $J_e$ and $J_{es}$. If
$J_e$ and $J_{es}$ are equal to the surface
coupling, $\beta_2$ (=$0.095 \pm 0.005$) is significantly
lower than the critical exponent of the surface
magnetization, $\beta_1= 1/8$. When weakening the
couplings at the edge, $\beta_2$ increases
monotonically. Eventually, $\beta_2$ will be clearly 
larger than 1/8, in accordance with related exact results for
two--dimensional Ising models with defect lines. The decrease
of $J_{es}$ has a more pronounced impact on $\beta_2$; the
value of the edge critical exponent may exceed appreciably
the limiting value obtained at vanishing $J_e$. By weakening
$J_e$, strong corrections to scaling for the edge magnetization
are observed.

It may be of interest to check the present Monte Carlo results by 
applying other methods, such as 
renormalization group techniques. Likewise, it may be feasible to
mimic the edge by a realistic ladder--type defect in a, possibly,
exactly solvable two--dimensional Ising model. Finally, the
non--universality of the critical behavior is expected to show
up in additional dependences of the critical exponents on details
of, for instance, the lattice structure
and the range of interactions.\\ 
 
\noindent {\bf Acknowledgement}
 
We thank F. Igl\'{o}i for a useful discussion.

\begin{figure}
\centerline{\psfig{figure=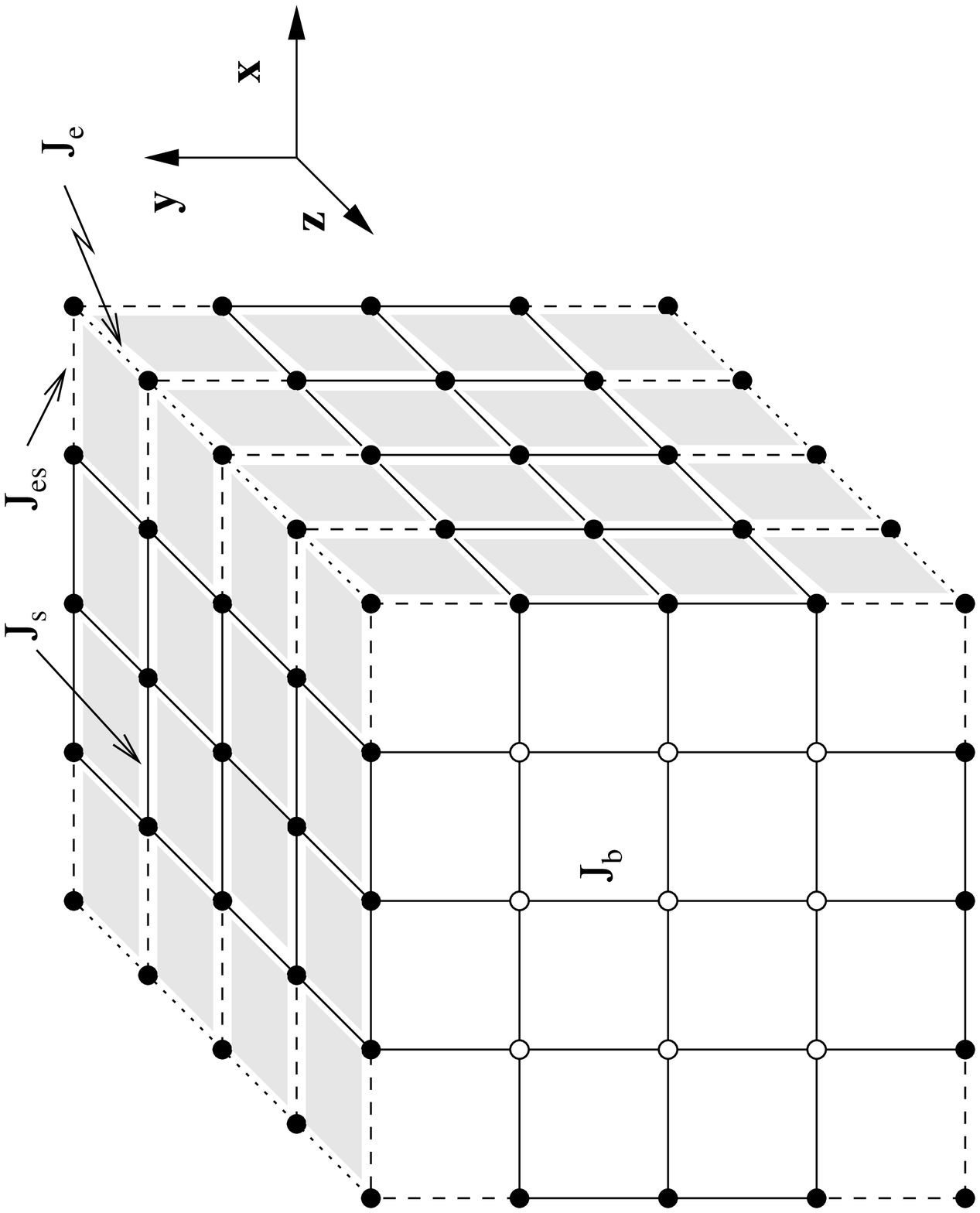,width=8.5cm,angle=270}}
\vspace*{0.2cm}
\caption{Sketch of the lattice structure and the coupling constants
at the edges, $J_e$ and $J_{es}$, at the surface, $J_s$, and
in the bulk, $J_b$.}
\label{fig1} \end{figure}
 
\begin{figure}
\centerline{\psfig{figure=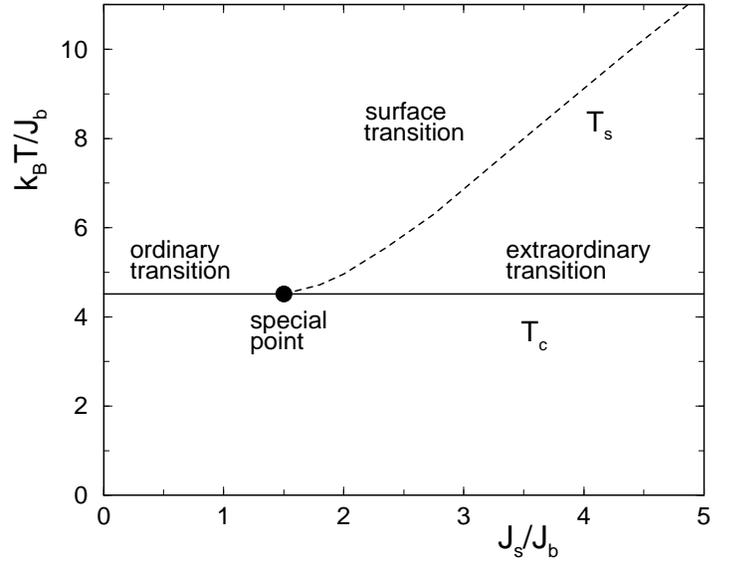,width=9.5cm,angle=270}}
\vspace*{0.2cm}
\caption{Phase diagram of the three--dimensional Ising model, in
the ($k_BT/J_b,J_s/J_b$)--plane. The dashed curve denotes the 
surface transition line, $T_s$.}
\label{fig2}
\end{figure}
 
\begin{figure}
\centerline{\psfig{figure=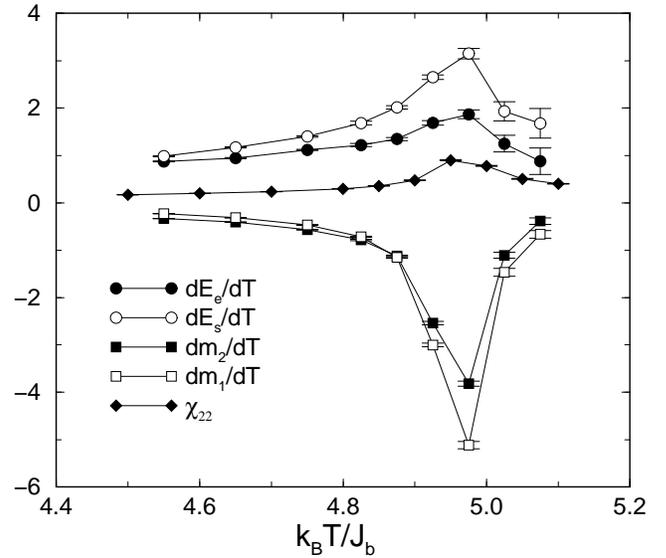,width=8.5cm,angle=270}}
\vspace*{0.2cm}
\caption{Monte Carlo data on temperature dependence of the  
edge susceptibility, $\chi_{22}$, and of the temperature derivatives of the
edge energy, $E_e$, surface energy, $E_s$, surface
magnetization, $m_1$, as well as  edge magnetization, $m_2$, near
$T_s$, at $J_e= J_{es}= J_s= 2J_b$ for a system of
size $40^2 \times 160$.}
\label{fig3}
\end{figure}
 
\begin{figure}
\centerline{\psfig{figure=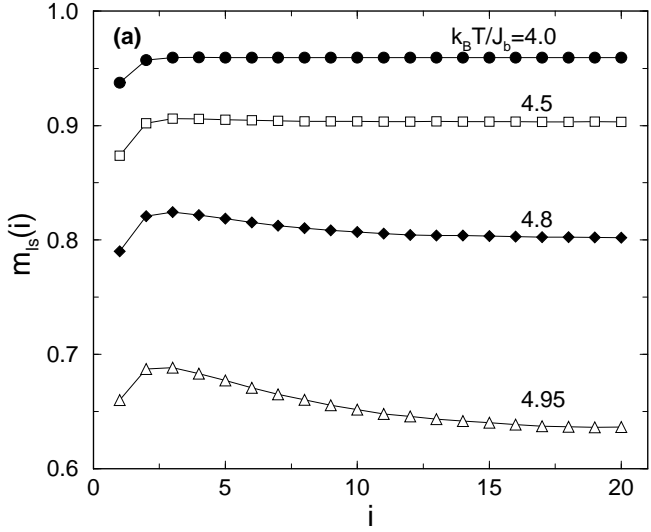,width=8.5cm,angle=270}}
\vspace*{0.5cm}
\centerline{\psfig{figure=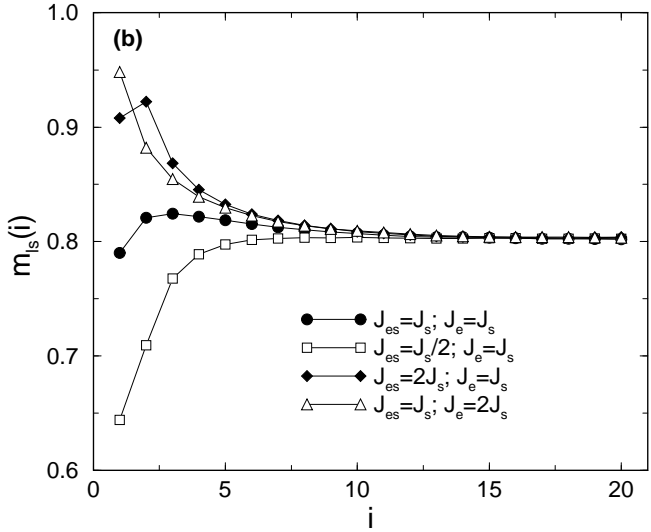,width=8.5cm,angle=270}}
\caption{Profiles of the line magnetization
$m_{ls}(i)$ at (a) $J_e= J_{es}= J_s= 2J_b$, with various temperatures, and
(b) at fixed temperature, $k_BT/J_b =4.8$, with various edge and edge--surface
couplings ($J_s= 2J_b$). Ising models of
size $40^3$ have been simulated.}
\label{fig4}
\end{figure}
 
\begin{figure}
\centerline{\psfig{figure=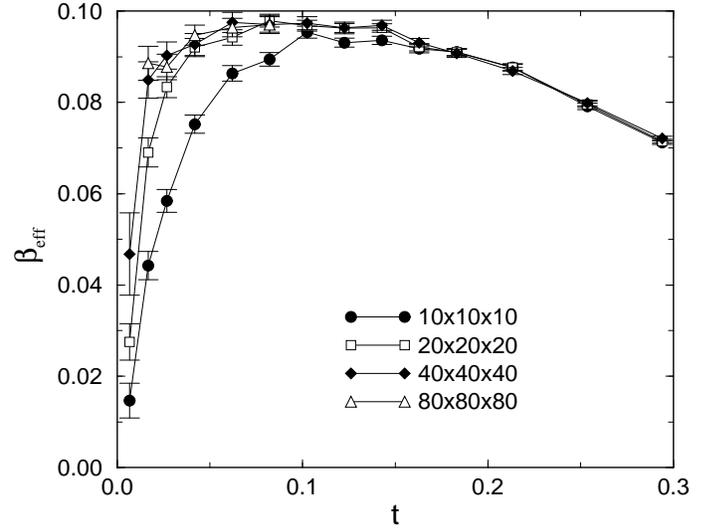,width=9.0cm,angle=270}}
\vspace*{0.2cm}
\caption{Effective exponent $\beta_{eff}$ of the edge magnetization 
as a function of reduced temperature $t= |T_s -T|/T_s$ at
$J_e= J_{es}= J_s= 2J_b$ for Monte Carlo systems of various sizes.}
\label{fig5}
\end{figure}
 
\begin{figure}
\centerline{\psfig{figure=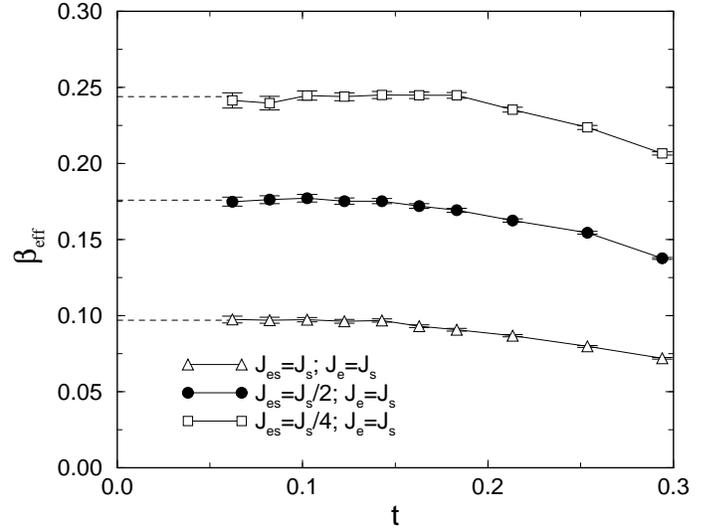,width=9.0cm,angle=270}}
\vspace*{0.2cm}
\caption{Effective exponent $\beta_{eff}$ versus 
reduced temperature $t$ for different edge--surface
couplings $J_{es}$, with $J_e= J_s= 2J_b$. System sizes have been adjusted
to circumvent finite--size dependences, with up to $80^3$ spins.}
\label{fig6}
\end{figure}
 
\begin{figure}
\centerline{\psfig{figure=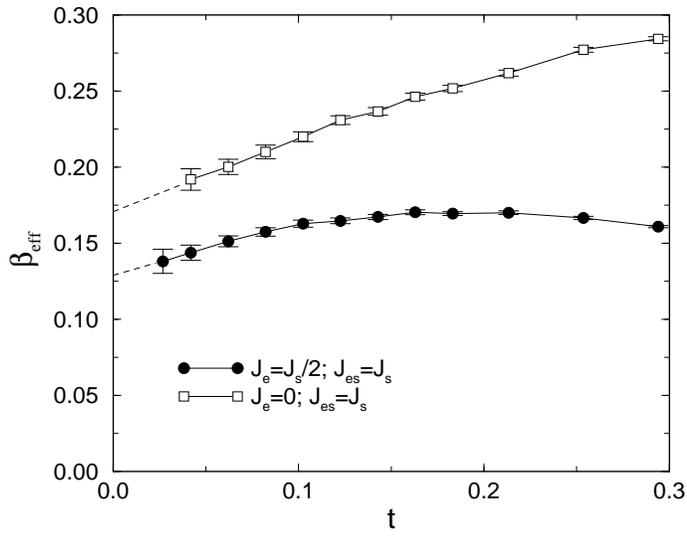,width=9.0cm,angle=270}}
\vspace*{0.2cm}
\caption{Effective exponent $\beta_{eff}$ versus
reduced temperature $t$ for different edge couplings
$J_e$, with $J_{es}= J_s= 2J_b$. System sizes for the simulations have 
been chosen to avoid finite--size effects, with up to $80^3$ spins close
to the surface transition.}
\label{fig7}
\end{figure}

\end{document}